\documentclass[letters,usenatbib]{mnras}

\usepackage{newtxtext,newtxmath}

\usepackage[T1]{fontenc}

\DeclareRobustCommand{\VAN}[3]{#2}
\let\VANthebibliography\thebibliography
\def\thebibliography{\DeclareRobustCommand{\VAN}[3]{##3}\VANthebibliography}

\usepackage{graphicx}	
\usepackage{amsmath}	
\usepackage{multirow}	

\newcommand{\kms}{\,km\,s$^{-1}$}
\newcommand{\msol}{\,M$_{\odot}$}

\newcommand{\um}{\,$\upmu$m}
\newcommand{\lsol}{\,L$_{\odot}$}
\newcommand{\sdv}{\,Jy\,km s$^{-1}$\,beam$^{-1}$}
\newcommand{\lpsol}{\,K\,km\,s$^{-1}$\,pc$^{2}$}

\title[A companion of the Cloverleaf]{Luck of the Irish? A companion of the Cloverleaf connected by a bridge of molecular gas}

\author[H.\,R.\,Stacey \& F.\,Arrigoni Battaia]{
H.\,R.\,Stacey$^{1}$\thanks{E-mail: stacey@mpa-garching.mpg.de}
and F.\,Arrigoni Battaia$^{1}$\\
$^{1}$Max Planck Institute for Astrophysics, Karl-Schwarzschild Str. 1, D-85748 Garching bei M\"unchen, Germany \\
}

\date{Accepted 2022 September 1. Received 2022 August 31; in original form 2022 July 27}

\pubyear{2022}

\begin{document}
\label{firstpage}
\pagerange{\pageref{firstpage}--\pageref{lastpage}}
\maketitle

\begin{abstract}
We present deep observations of CO\,(3--2) from the Cloverleaf lensed quasar-starburst at $z=2.56$. We discover a 4--5 times less massive companion at a projected distance of 33~kpc from the Cloverleaf host galaxy. The galaxies are connected by a bridge of CO emission, indicating that they are interacting and that the companion is being stripped by the Cloverleaf. We also find evidence for fast molecular gas in the spectral line of the Cloverleaf that may be an outflow induced by stellar or quasar feedback. All of these features may be ubiquitous among quasars and only detected here with the help of gravitational lensing and the sensitivity of the data. Overall, these findings agree with galaxy formation scenarios that predict gas-rich mergers play a key role in quasar triggering, starburst triggering and the formation of compact spheroids.
\end{abstract}




\section{Introduction}
\label{sec:intro} 

Hierarchical galaxy formation models predict massive elliptical galaxies formed as a result of gas-rich mergers \citep{Kauffmann:2000,Hopkins:2008a,Hopkins:2008b}. The very high central stellar densities of elliptical galaxies can be produced if gas is funnelled into the centre of the galaxy faster than the star formation rate, generating a compact starburst and rapid stellar growth \citep{Mihos:1996,Valentino:2020}. At the same time, rapid accretion of gas onto the central supermassive black hole creates an active galactic nucleus (AGN) that injects energy and momentum into the interstellar medium in the form of radiation pressure \citep{Fabian:2008}, relativistic winds \citep{King:2010} or radio jets \citep{Mukherjee:2018}. These processes can explain the scaling relations between galaxy properties which seem to already be in place by $z\sim2$ \citep{ForsterSchreiber:2020}.

Hierarchical models naturally predict that quasars (the most energetic AGN) have many companion galaxies, the number and properties of which can probe the strength of feedback on the surrounding environment \citep{Costa:2019}. Previous studies have found that some quasars at $z>2$ have dust-obscured companions \citep{Decarli:2017,Trakhtenbrot:2017,Banerji:2018,Neeleman:2019,Chen:2021,Bischetti:2021}. Sub-mm observations of dust-continuum of $z\sim2$ quasars suggests that they reside in over-densities, although most studies lack spectroscopic information to confirm a physical association \citep{Silva:2015,Hatziminaoglou:2018}. Depending on the properties of the companions, they may be faint in UV and/or the sub-mm emission, and only detected in spectral line emission \citep{Drake:2020}.

If quasars are triggered as a result of mergers, we should expect to see the imprint of the galaxy interactions on the cold interstellar medium of their host galaxies. \cite{Neeleman:2019} and \cite{Drake:2020} found clear evidence for this by detecting diffuse [CII] emission between a quasar and its dusty starburst companion at $z>4$. Simulations suggest we should also expect to observe the effect of AGN feedback in the form of multi-phase outflows \citep{Costa:2020}. The molecular gas kinematics of quasar host galaxies have not been studied in detail at $z\sim2\--4$ beyond resolving apparently coherent velocity structure \citep{Stacey:2021,Banerji:2021}, although \cite{Feruglio:2017} find evidence for a component of fast molecular gas in the CO\,(4--3) line emission from gravitationally lensed hyper-luminous quasar-starburst APM\,08279+5255 at $z=4$, and \cite{Spingola:2020} found disturbed CO\,(1--0) gas kinematics for two lensed quasar hosts at $z=2$ and 3. Furthermore, \cite{Stacey:2022} detect quasar-driven molecular outflows from dust-reddened quasars, but not unobscured quasars, suggesting that they may only be detected in a molecular phase for a fraction of the quasar lifetime. 

We have tested these predictions by analysing deep CO\,(3--2) observations of the Cloverleaf (H\,1413$+$117), a quasar at $z=2.56$ with an obscured host galaxy that is gravitationally lensed by an intervening massive elliptical galaxy \citep{Magain:1988,Riechers:2011b,Stacey:2021}. The galaxy contains an Eddington-limited starburst and the strong X-ray radiation field may contribute to the molecular gas heating \citep{Bradford:2009,Riechers:2011a,Uzgil:2016}. In Section~\ref{section:data}, we discuss the observations and data reduction. In Section~\ref{section:results}, we detail our analysis of the line profile and spectral cube. Finally, in Section~\ref{section:discussion}, we consider implications for hierarchical formation models and future work.

\section{Data}
\label{section:data} 

The Cloverleaf was observed with ALMA under project code 2017.1.01232.S (PI: Nishimura) in 9 epochs on 21 Dec 2017, 29 Dec 2017, 9 Jan 2018 and 11 Jan 2018. The observations were taken in an antenna configuration with maximum baseline 2.5 km. The data were correlated in both linear polarisations (XX,YY) in four spectral windows, with one spectral window centred on the redshifted rest-frequency of the CO\,(3--2) line (97.19267587~GHz).

The calibrated visibilities were produced using the ALMA pipeline in CASA \citep{McMullin:2007}. The data were inspected to confirm the quality of the calibration and that no further flagging was required. The spectral line data were prepared by subtracting the continuum from the visibility data using the task $uvcontsub$ to fit a model to the line-free spectral windows. The line profile of the Cloverleaf and its companion are shown in Fig.~\ref{fig:lineprofile}, each extracted from a circular aperture around the object to maximise signal-to-noise. Moment maps of the deconvolved image cubes are shown in Fig.~\ref{fig:moments} with natural weighting of the visibilities. We measure an off-source rms noise of $1\times10^{-4}~{\rm Jy\,km\,s^{-1}\,beam^{-1}}$ in the image cube for a spectral resolution of 15\kms\ with natural weighting. The line emission integrated around the line velocity of the companion with $uv$-taper equivalent to 0.7~arcsec is shown in Fig.~\ref{fig:bridge}. 

\begin{figure}
    \centering
    \includegraphics[width=0.43\textwidth]{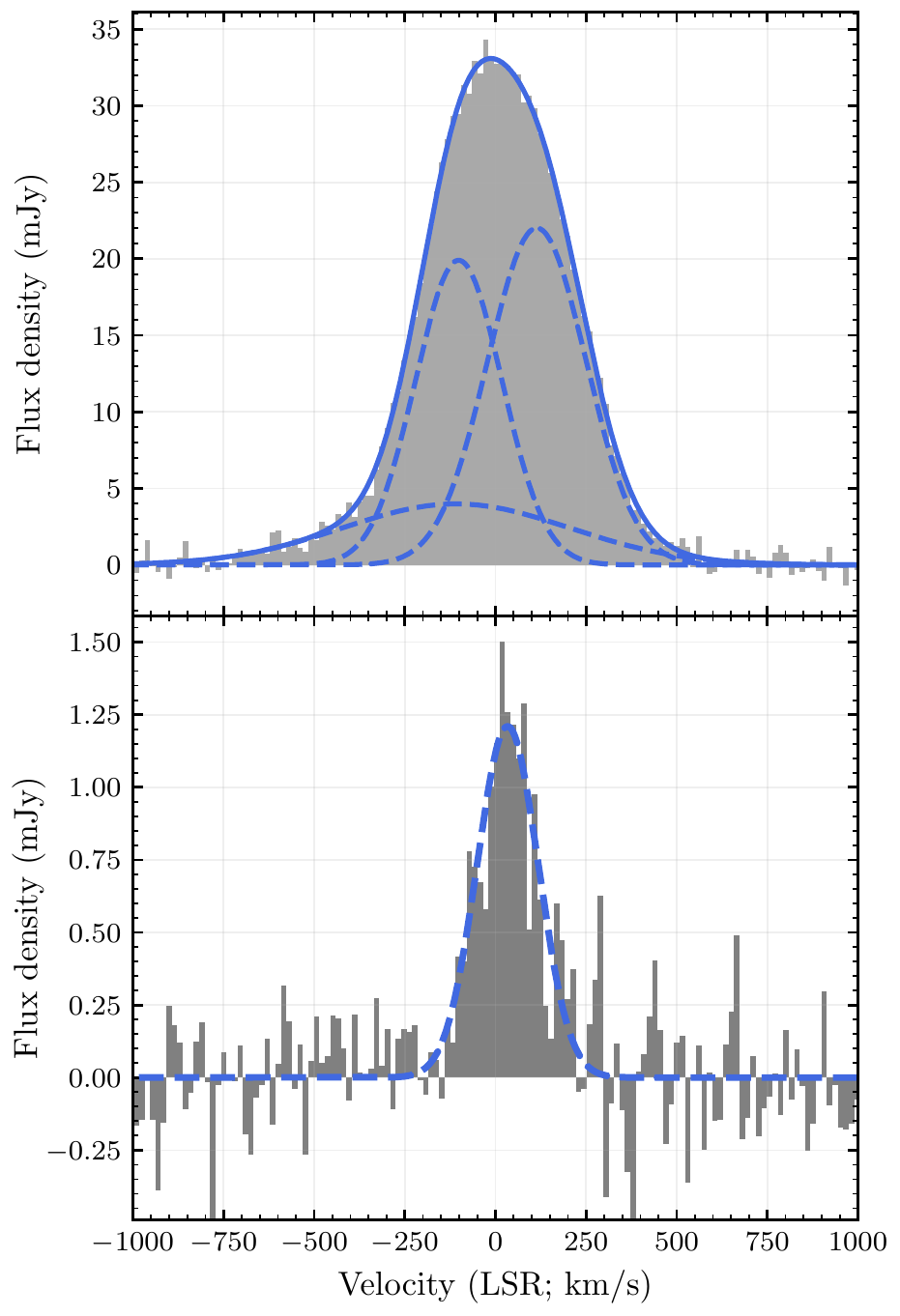}
    \caption{Line profiles of the Cloverleaf (top) and the companion (bottom) with 15\kms\ spectral resolution. The spectral line emission from Cloverleaf is fit by three Gaussian components, including a broad line of FWHM $740\pm70$\kms. The assumed systemic redshift is $z=2.55784$ from \citep{Stacey:2021}.}
    \label{fig:lineprofile}
\end{figure}

\begin{figure*}
    \centering
    \includegraphics[width=\textwidth]{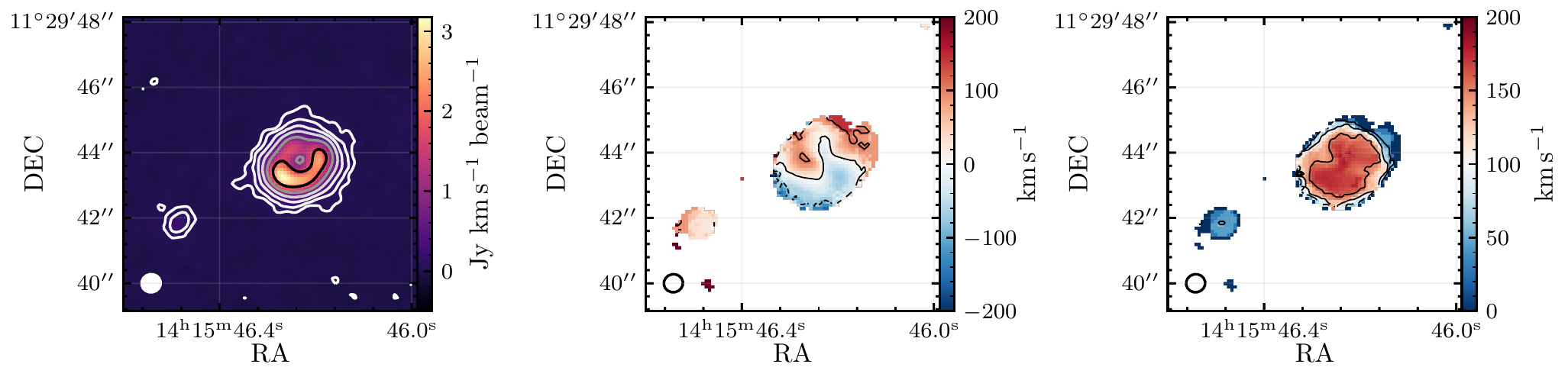}
    \includegraphics[width=\textwidth]{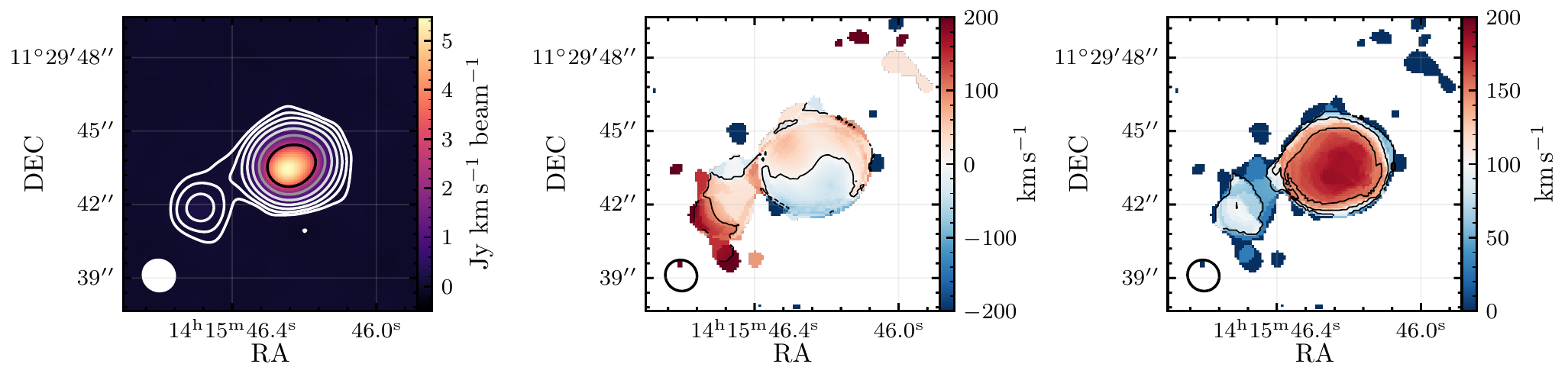}
    \vspace{-0.5cm}
    \caption{Moment maps of CO\,(3--2) emission from the Cloverleaf and its neighbour (four arcsec to the South-East). Top row is naturally weighted imaging (0.5~arcsec resolution); bottom row is with a $uv$ taper equivalent to 1~arcsec (1.2~arcsec angular resolution). Panels show velocity integrated line intensity, velocity field, velocity dispersion. The contours on the moment 0 image are signal-to-noise of $-3$, 3, 6, 12, 24.. etc. The contours in the moment 1 and 2 are steps of 100\kms. The synthesised beam effective FWHM is shown in the bottom-left corner. }
    \label{fig:moments}
\end{figure*}

\begin{figure}
    \centering
    \includegraphics[height=0.3\textwidth]{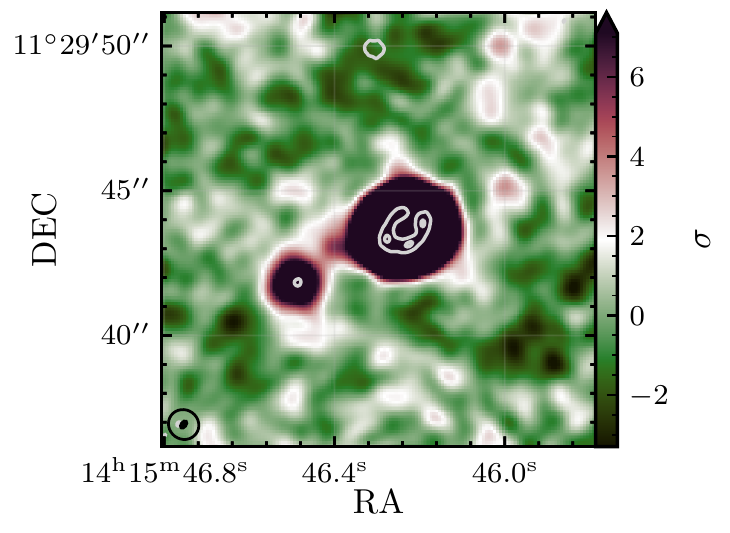}
    \vspace{-0.2cm}
    \caption{The CO\,(3--2) line emission at 1~arcsec angular resolution, integrated over the velocity channels of the companion line profile (Fig.~\ref{fig:lineprofile}) and normalised to the rms noise. The contours are 334~GHz continuum surface brightness at 0.3 and 3~mJy\,beam$^{-1}$, where the third object (6~arcsec North) is also seen. The synthesised beams are shown in the lower-left corner.}
    \label{fig:bridge}
\end{figure}

\begin{table*}
    \centering
    \caption{CO line properties for the two sources. For the calculation of the molecular gas mass of the Cloverleaf, we assume the average magnification of the CO\,(3--2) is $\mu=11$ as was found for CO\,(9--8) reported by \protect\cite{Stacey:2021}. All other given properties are uncorrected for lensing magnification. The total line emission is measured directly from the moment 0 image (hence, the smaller uncertainties). The central velocity of the Gaussian components are relative to the systemic velocity reported by \protect\cite{Stacey:2021}.}
    \begin{tabular}{l l c c c c c c} \hline
                    & & $v_{\rm CO}$ & FWHM & $I_{\rm CO}$ & $\mu L_{\rm CO}$ & $\mu L^{'}_{\rm CO}$ & $M_{\rm mol}$ \\ 
                    & & (\kms) &  (\kms) & (\sdv)     & (\lsol)      & (\lpsol)          & (\msol) \\ \hline
       \multirow{3}{*}{Cloverleaf} &component 1 & $110\pm20$ & $320\pm30$ & $7.5\pm1.2$ & $3.5(\pm0.6)\times10^8$ & $2.6(\pm0.4)\times10^{11}$ & $1.9(\pm0.3)\times10^{10}$ \\
                                & component 2 & $-100\pm20$ & $280\pm20$ & $5.9\pm1.3$ & $2.8(\pm0.6)\times10^8$ & $2.1(\pm0.5)\times10^{11}$ & $1.5(\pm0.3)\times10^{10}$ \\
                                & component 3 & $-110\pm40$ & $740\pm70$ & $3.1\pm0.8$ & $1.5(\pm0.4)\times10^8$ & $1.1(\pm0.3)\times10^{11}$ & $0.8(\pm0.2)\times10^{10}$ \\
                                & total       & &  & $15.8\pm0.1$ & $7.4(\pm0.1)\times10^8$ & $5.0(\pm0.1)\times10^{11}$ & $4.0(\pm0.1)\times10^{10}$ \\ 
    Companion                   & &   $34\pm6$  & $200\pm15$        & $0.31\pm0.04$ & $1.5(\pm0.2)\times10^7$           & $1.1(\pm0.1)\times10^{10}$ & $0.9(\pm0.1)\times10^{10}$  \\ \hline
    \end{tabular}
    \label{tab:line_properties}
\end{table*}

\section{Results}
\label{section:results} 

We clearly detect a companion in the CO\,(3--2) line emission at a distance of 4~arcsec ($\approx33$~kpc) South-East of the Cloverleaf (J2000 RA: 14:15:46.49, Dec: 11.29.41.8).  The gravitational lensing magnification at this projected distance is negligible. The companion is close to the systemic velocity of the Cloverleaf in CO\,(3--2) and CO\,(9--8) line emission ($z=2.558$; \citealt{Stacey:2021}), slightly redshifted by $34\pm6$\kms, although this is within uncertainties of the Cloverleaf systemic velocity. The companion has an integrated line intensity of $0.31\pm0.04$\sdv, relating to a luminosity of $L'_{\rm CO}=1.1(\pm0.1)\times10^{10}~{\rm K\ km\ s^{-1}\ pc^2}$. Assuming a conversion factor of $0.8\,{\rm (K\ km\ s^{-1}\ pc^2)^{-1}}$, typically adopted for high-redshift starbursts \citep{Greve:2014}, this companion galaxy has a molecular gas reservoir of $0.9(\pm0.1)\times10^9$\msol. The galaxy is marginally resolved in the direction of the Cloverleaf.

In contrast, the Cloverleaf has an integrated line luminosity of $15.8(\pm0.1)$\sdv. Assuming a magnification of $\approx11$, as found for both the CO\,(9--8) and sub-mm dust emission by \cite{Stacey:2021}, we infer a molecular gas reservoir of $4.0(\pm0.1)\times10^{10}$\msol, suggesting that the companion is 4--5 times less massive than the Cloverleaf from the perspective of its molecular gas content (Table~\ref{tab:line_properties}). The molecular gas mass we infer for the Cloverleaf is larger than the $3.1\times10^{10}$\msol\ found by \cite{Riechers:2011b} with large-velocity gradient modelling of the CO spectral line energy distribution (although no uncertainties were given) and consistent with the 0.2--$5\times10^{10}$\msol\ inferred by \cite{Bradford:2009} with a similar approach.

We searched for evidence of the companion in archival ALMA observations of the Cloverleaf and detect it at $0.5\pm0.1$~mJy in 334~GHz (rest-frame 250\um) continuum (2012.1.00175.S; PI: van der Werf) where it is unresolved with $\approx0.2$~arcsec angular resolution. The companion is about 6 times fainter than the intrinsic flux density of the Cloverleaf at similar frequencies. We did not detect the companion in other CO lines, H$_2$O or [CI]\,(1--0), probably due to the sensitivity of the observations. Although we note that there is a slight extension towards the location of the companion in low-angular-resolution observations of [CI]\,(2--1) by \cite{Weiss:2003}. We did not find any association for the companion in R, I or H-band HST imaging, suggesting that the companion is also dust-obscured.

We resolve an extension of molecular gas connecting the Cloverleaf and its companion at the systemic velocities of the companion. This can be seen clearly in the moment maps where we have applied a $uv$-taper to increase the surface brightness sensitivity (Fig.~\ref{fig:bridge}). This intergalactic bridge indicates that gas is being stripped by ram pressure or a tidal interaction. The gas in the companion and in the bridge is close to the systemic velocity, suggesting that the direction of motion is perpendicular to the line-of-sight.

Another neighbouring galaxy was detected 6~arcsec North of the Cloverleaf in dust emission with ALMA by \cite{Stacey:2021} (see also Fig.~\ref{fig:bridge}). No spectral line emission was detected in either CO\,(9--8) or CO\,(3--2) in any spectral window, although it is much brighter in continuum emission than the close companion, suggesting that this third object is not physically associated with the Cloverleaf. We searched for other close companions to the Cloverleaf within $\pm1000$\kms\ within the primary beam ($\approx500$~kpc radius) using the publicly available code LineSeeker \citep{Gonzalez-Lopez:2017}, requiring any detection to have a FWHM $<1000$\kms\ and a signal-to-noise of $>3$ in three adjacent channels (equivalent to $L'_{\rm CO}\geq10^{8.9}$\lpsol). We did not find evidence for additional companions.

\begin{figure}
    \centering
    \includegraphics[height=0.3\textwidth]{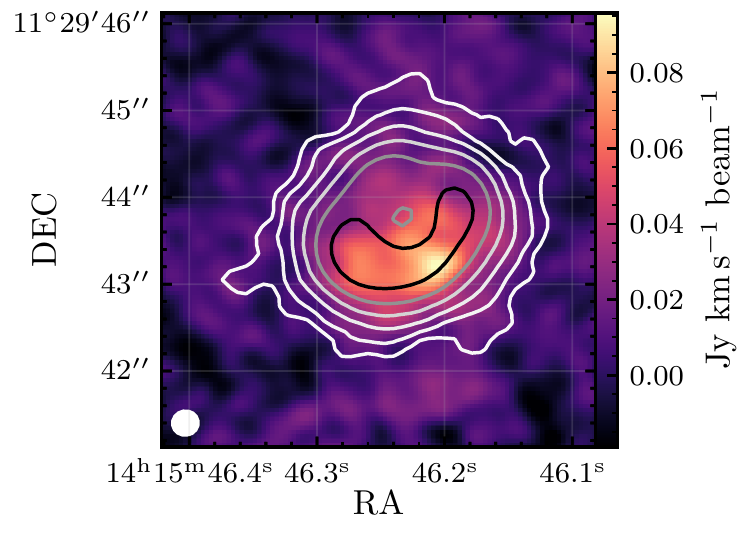}
    \vspace{-0.2cm}
    \caption{The CO\,(3--2) line emission from the Cloverleaf integrated between $-750$ and $-400$\kms with natural weighting. The contours are the total integrated line emission, as in Fig.~\ref{fig:moments}.}
    \label{fig:outflow}
\end{figure}

The best fit to the line profile from the Cloverleaf comprises three Gaussian components, two of which could be consistent with gas in a disc, and an additional broad component with a FWHM of $740\pm70$\kms. The integrated line emission between $-750$ and $-450$\kms\ is show in Fig.~\ref{fig:outflow}. The location of this excess blue emission suggests is located within the quasar host galaxy but not coincident with the quasar (see also Figs. 2d and 4d by \citealt{Stacey:2021}). This could be an outflow driven by star formation \citep{Ginolfi:2020} or quasar feedback \citep{Stacey:2022}. It may also be strongly differentially lensed. A similar feature was detected in the CO\,(4--3) line profile of the hyper-luminous strongly lensed quasar APM\,08279+5255 \citep{Feruglio:2017} where it is also unclear whether the feature is due to differential lensing. The Gaussian fit suggests that the mass in the possible outflow is about 20~percent of the molecular gas mass of the Cloverleaf and comparable to the gas mass of the companion (Table~\ref{tab:line_properties}), although we stress that this is highly dependent on the CO--H$_2$ conversion factor and local lensing magnification.

\section{Discussion and conclusions}
\label{section:discussion}

\subsection{The companion} 

Cosmological hydro-dynamical simulations and semi-analytic models predict that gas-rich mergers play a key role in forming elliptical galaxies by causing cold gas to lose angular momentum, triggering a compact starburst \citep{Bois:2011,Duc:2011,Naab:2014,Mihos:1996,Hopkins:2008a}. Observations of $z>2$ quasars are broadly consistent with this model, being located within galaxy over-densities \citep{Decarli:2017,Drake:2020} some of which have disturbed gas kinematics \citep{Litke:2019,Spingola:2020,Neeleman:2021}. Here, we find clear evidence for an ongoing interaction and displacement of cold gas in a quasar-starburst, in agreement with a merger-triggering scenario. Our results suggest Cloverleaf and its companion are likely hosted in a massive dark matter halo \citep{Kauffmann:2002,Costa:2014}. 

Surveys of quasar and blank fields do not reach the limiting CO\,(3--2) luminosity of these data \citep{Decarli:2017}. While not much can be inferred from our single companion detection due to Poisson uncertainties, we can expect statistical observations with similar quality to give strong constraints on quasar environments at $z\sim2\--3$.

\subsection{The bridge} 

An outflow is unlikely to explain the presence of the bridge that connects the Cloveleaf to its companion. As the velocities in the bridge are close to the systemic velocity of the galaxies with relatively small velocity dispersion (Fig.~\ref{fig:moments}), it is not consistent with what is typically observed for molecular outflows from quasars (e.g. \citealt{Feruglio:2010}). An outflow must have a very narrow opening angle that is coincidentally directed towards the companion, in the plane of the sky.

Another explanation for the presence of the intergalactic bridge is that gas in the less massive companion is being stripped during the interaction with the Cloverleaf or the intergalactic medium. Tidal interactions and ram-pressure stripping are potential mechanisms. Merger tails and ram-pressure-stripped gas have been found to be dynamically cold with velocities similar to the parent galaxy \citep{Hibbard:1996,Wild:2014}, but may be dynamically hot at larger distances ($>20$~kpc) where it transitions from a laminar to turbulent flow \citep{Fumagalli:2014}: the low velocity dispersion in the bridge suggests that the gas from the companion is dynamically cold. Following \cite{Mo:2010}, we estimate the tidal stripping radius assuming the impulse approximation to be $\sim 1$~kpc using the ratio of gas masses and projected distance from the Cloverleaf, suggesting that tidal stripping is ongoing. We see no evidence of an extended tidal tail on the opposite side of the companion that is expected due to conservation of angular momentum, although this could be because of its very low surface brightness. Adopting the formula to estimate ram-pressure stripping from \cite{Domainko:2006}, we compute the stripping radius assuming a perpendicular velocity equivalent to the relative systemic velocity of the companion, and a stellar mass of $6\times10^9$\msol\ computed from the difference between its dynamical mass (following \citealt{Neeleman:2021}) and gas mass. Assuming a halo density of $10^{-3}~{\rm cm^{-3}}$ from \cite{Nelson:2016}, we find a ram-pressure-stripping radius of $\sim 10$~kpc. This suggests ram pressure stripping is unlikely to be responsible for the molecular bridge unless the perpendicular velocity of the companion is $\sim10$ times larger\footnote{This would be a more realistic relative velocity and mean that most of the companion's motion is on the plane of the sky. But for a velocity of $\sim100$ times larger, the companion would not be gravitationally bound to the halo} {\it and} the halo density is $\gtrsim100$ times larger than found in hydrodynamical simulations. However, we stress that these are probably no better than order-of-magnitude estimates. Detection of the Sunyaev-Zeldovich effect \citep{Sunyaev:1972} and/or X-ray radiating hot gas could test whether a halo, with sufficiently high temperature and density for ram-pressure stripping to occur, is already established around the Cloverleaf.

Observations of tidal stripping in galaxy interactions in the local Universe suggest that the stripped gas originates in the parent galaxy \citep{Lisenfeld:2004,Lisenfeld:2008}. However, observations of low redshift galaxies that are being stripped by cluster environments and tidal interactions show evidence for molecular gas formation \citep{Moretti:2020} and star formation \citep{deMello:2008,Fumagalli:2011,ArrigoniBattaia:2012,Fossati:2016} in the stripped gas, which could also be responsible for the CO\,(3--2) emission.

\subsection{The fast molecular gas} 

The fast molecular gas in the Cloverleaf system could have a number of origins, including turbulent ram-pressure-stripped gas \citep{Fumagalli:2014}, feedback from star formation or AGN feedback. There are multiple AGN feedback mechanism that appear be ongoing within the galaxy: radiation pressure \citep{Stacey:2022}, relativistic winds \citep{Chartas:2007} and radio jets (Peters et al. in prep), as well as an Eddington-limited starburst \citep{Bradford:2009,Stacey:2021}, so both AGN and star formation feedback are highly plausible. Future work will involve detailed lensing-kinematic modelling (e.g. \citealt{Spingola:2020,Stacey:2021,Rizzo:2021}) to resolve the gas kinematics in the Cloverleaf host galaxy and diagnose the nature of the high-velocity molecular gas.

Finally, we emphasise that discoveries reported in this letter are only possible due to the combination of gravitational lensing and deep ALMA observations. If the Cloverleaf were not lensed, it would have to be observed for $\approx120$ times as long ($\approx1000$~hrs) to achieve comparable signal-to-noise. Therefore, the fast molecular gas (and perhaps the intergalactic bridge) may be ubiquitous among quasars at cosmic noon and beyond, but undetectable with current instruments without the aid of gravitational lensing. This also raises the possibility that the extended haloes of cold gas detected around quasars $z>4$ from stacking \citep{Bischetti:2019,Stanley:2019} are actually detecting these intergalactic structures and/or companion galaxies. Observations of gravitationally lensed quasars with comparable quality will be required to test whether the properties of the Cloverleaf are typical (or if we have been lucky).

\section*{Acknowledgements}

We thank Ryan Farber and Adam Schaefer for helpful discussions. Our research used Astropy, NumPy and Matplotlib packages for Python \citep{AstropyCollaboration:2013,AstropyCollaboration:2018,Harris:2020,Hunter:2007}. HRS acknowledges funding from the European Research Council (ERC) under the European Union's Horizon 2020 research and innovation programme (LEDA: grant agreement No 758853). We made use of ALMA data with project codes 2017.1.01232.S and 2012.1.00175.S. ALMA is a partnership of ESO (representing its member states), NSF (USA) and NINS (Japan), together with NRC (Canada), MOST and ASIAA (Taiwan), and KASI (Republic of Korea), in cooperation with the Republic of Chile. The Joint ALMA Observatory is operated by ESO, AUI/NRAO and NAOJ. 

\section*{Data Availability}

All observations reported in this work are publicly available in the ALMA archive (\url{https://almascience.eso.org/aq}) and all the analysis was performed with publicly available software. The data sets generated during this study are available from the corresponding author upon reasonable request.

\bibliographystyle{mnras}
\bibliography{references} 

\bsp	
\label{lastpage}
\end{document}